\documentclass[12pt]{iopart}
\usepackage{epsf,psfig}

\begin{document}

\title{Salient Features of Hadronization, Deconfinement and Heavy Quark Probes at RHIC}

\author{Huan Z. Huang\dag}

\address{Department of Physics and Astronomy, University of California, Los Angeles, CA 90095-1547, USA}

\address{\dag\ Email: huang@physics.ucla.edu}

\begin{abstract}
We present experimental features of identified particle production from nucleus-nucleus collisions at RHIC. These features 
reflect hadronization from a deconfined partonic matter whose particle formation scheme is distinctly different
from fragmentation phenomenology in elementary collisions. Multi-parton dynamics, such as quark coalescences or 
recombinations, appear to be essential to explain the experimental measurements at the intermediate transverse 
momentum of 2-5 GeV/c. Constituent quarks seem to 
be the dominant degrees of freedom at hadronization and gluon degrees of freedom are not explicitly manifested in the 
hadronization scheme. This physical scenario is consistent with recent Lattice QCD calculations that near the critical 
temperature hadrons do not melt completely and quasi-particles may provide an effective degrees of freedom to 
describe the partonic 
matter. Heavy quark production and other future experimental measurements to quantify 
the QCD properties of the produced matter at RHIC will be discussed.

\end{abstract}

\section{Introduction}

The advent of the Relativistic Heavy Ion Collider (RHIC) at Brookhaven
National Laboratory (BNL) has ushered in a new area in searches for the
Quark Gluon Plasma (QGP). Recent measurements by all four RHIC experiments
have established that a dense medium has been created in central Au+Au
collisions at RHIC and the energy loss of partons leads to a suppression of
high transverse momentum ($p_{T}$) particles~\cite{dAu-star,dAu-phenix,dAu-phobos,dAu-bhrams}. 
The QCD nature of the dense
matter created at RHIC and whether the current experimental evidence proves
the discovery of the QGP have been under debate within the heavy ion physics
community~\cite{RBRC-1,RBRC-2,RBRC-3}. We will attempt to address three specific questions: 
1) features of hadronization and evidence for a color
deconfined bulk partonic matter; 2) QCD properties of the matter at
hadronization from experimental data and from Lattice QCD calculations; and
3) heavy quark production and other future experimental measurements which would further
quantify the QCD properties at the phase boundary. 

\section{Features of Hadronization From Bulk Matter}

QCD color charges are confined and hadrons exist in color singlet state. The
hadronization of quarks and gluons (partons) has not been understood based
on QCD principles. Phenomenologically hadron formation in elementary 
$e^{+}e^{-}$ and nucleon-nucleon collisions has been described by
fragmentation processes. In the pQCD domain physical processes are
factorized into parton distribution functions, parton interaction processes
and fragmentation functions for hadron formation. The fragmentation function
is assumed to be universal and can be obtained from $e^{+}e^{-}$ collisions.
A typical Feynmen-Field~\cite{FF} fragmentation process involves a leading parton
of momentum $p$, which fragments into a hadron of momentum $p_{h}$ whose
properties are mostly determined by the leading parton. The fragmentation
function is a function of variable $z=p_{h}/p,$ where $z$ is between 0 and
1. Baryon production is found to be significantly suppressed compared to the production of
light mesons: the baryon to pion ratio increases with $z$, but never
exceeds $20\%$~\cite{SLD}.

In the soft (low $p_{T}$) particle production region, the pQCD framework and
factorization break down -- particles are believed not to be from
fragmentation of partons. String fragmentation models were inspired by the
QCD description of quark and anti-quark interaction. 
The Lund string model is one of the popular hadron formation
models which has been successfully implemented in Monte Carlo description of
e+e, p+p and nuclear collisions~\cite{LUND}. In string fragmentation models the baryon production
is also suppressed because the baryon formation requires the clustering of
three quarks~\cite{lund-b}.

Baryon production from nucleus-nucleus collisions, especially for
multi-strange hyperons, has been measured to be much larger than theoretical
model calculations. The hyperon production per number of participant pairs
from nucleus-nucleus collisions at the SPS is significantly enhanced in
comparison with the value from p+p collisions. The enhancement factor
increases from $\Lambda $ to $\Omega $ hyperons and with the collision
centrality. The large production of strange hyperons cannot be described by
string fragmentation models. In nucleus-nucleus collisions a large number of
strings would overlap and lead to a much stronger string tension.
A theoretical model to describe multiple strings as a rope has been proposed~\cite{Rope}. 
Strange hyperon production would be increased from ropes or strings with
a larger tension. 
The large enhancement in the strange baryon production with respect to the
number of participant scaling from p+p collisions has been proposed as
evidence for the formation of the QGP at the SPS~\cite{LRT}.
However, there is no direct experimental measurement linking the hyperon
production with the QGP hadronization. Furthermore, theoretical calculations
based on chiral properties of hadrons at high density~\cite{Soff} or strong color 
fields~\cite{CF} can also provide
an explanation for the large hyperon production.

\begin{figure}[htb]
   \centering
   \epsfxsize=8 cm
   \epsfysize=8 cm
   \leavevmode
   \epsffile{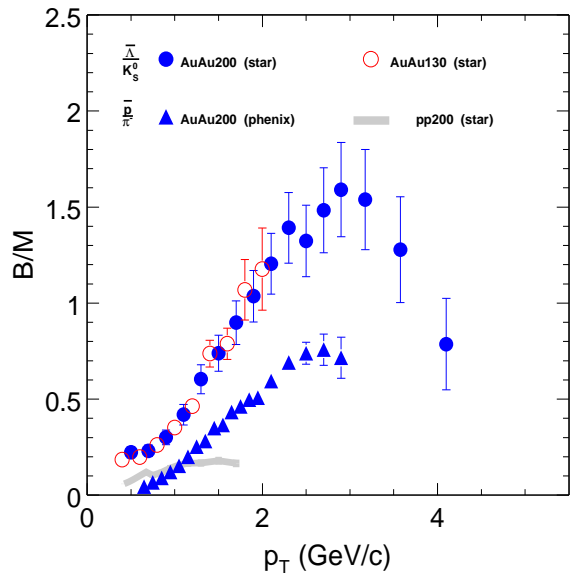} 
\caption[ratio]{Ratios of ${\overline{\Lambda}}$ to $K_{S}$ from Au+Au and p+p collisions (STAR) and 
$\overline{p}$ to $\pi$ from Au+Au collisions (PHENIX) as a function of transverse momentum ($p_{T}$).
The $\overline{\Lambda}$ data include contributions from $\Sigma^{0}$ decays and the $\overline{p}$ data
are corrected for weak decays of hyperons.}
\label{ratio}
\end{figure}

The increase in the baryon production from nucleus-nucleus collisions has become
much more prominent at RHIC energies. Figure~\ref{ratio} shows the ratios of $\overline{p}/\pi $
and $\overline{\Lambda }/K_{S}$ from central Au+Au collisions at 
$\sqrt{s_{NN}}=130,200$ GeV measured by PHENIX~\cite{phenix-b1} and STAR~\cite{star-b1,star-b2,star-k}. The apparent difference
between the STAR and PHENIX ratios can be explained by the fact that the
STAR $\overline{\Lambda }$ data include the electromagnetic decay contribution
from $\Sigma^{0}$ and the PHENIX data are corrected for weak decay
contributions. The STAR $\overline{\Lambda}$ data have been corrected for feeddown contributions from multi-strange hyperon decays.
Some early data, for example~\cite{phenix-b2}, were not included because these 
data were not corrected for weak decay contributions.
The large baryon to meson ratio cannot be accommodated by the
fragmentation scheme. The large ratio at the intermediate $p_{T}$ region is
the first indication that particle formation dynamics in nucleus-nucleus
collisions at RHIC are distinctly different from the hadron formation
mechanism via fragmentation processes in elementary $e^{+}e^{-}$ and
nucleon-nucleon collisions.

Measurements on nuclear modification factors and azimuthal angular
anisotropy for identified particles have provided essential insight into
hadronization in nucleus-nucleus collisions. The nuclear modification factor
is defined as 
\begin{equation*}
R_{AA}=\frac{[yield]^{AA}}{N_{bin}\times \lbrack yield]^{pp}}
\end{equation*}
where $N_{bin}$ is the number of binary nucleon-nucleon collisions. The 
$[yield]^{AA}$ and $[yield]^{pp}$ are particle yields ($d^{2}n/dp_{T}dy$)
from A+A and p+p collisions, respectively. The nuclear modification factor
has also been defined using peripheral and central collisions 
\begin{equation*}
R_{CP}=\frac{[yield/N_{bin}]^{central}}{[yield/N_{bin}]^{peripheral}}.
\end{equation*}
A $R_{AA}$ or $R_{CP}$ of unity implies that particle production from
nucleus-nucleus collision is equivalent to a superposition of independent
nucleon-nucleon collisions. Hard scattering processes within the pQCD
framework are believed to follow approximately binary scaling in the
kinematic region where the nuclear shadowing of parton distribution function
and the Cronin effect are not significant. Measurements of charged hadrons and neutral pions have
revealed a strong suppression at high $p_{T}$ region in central
Au+Au collisions~\cite{highpt-star1,highpt-star2,highpt-phenix1,highpt-phenix2}. 
Recent d+Au measurements~\cite{dAu-star,dAu-phenix,dAu-phobos,dAu-bhrams} 
have demonstrated that the large
suppression of high $p_{T}$ particles in central Au+Au collisions is mainly
due to energy loss, presumably of partons traversing the dense matter created
in these collisions.

\begin{figure}[htb]
   \centering
   \epsfxsize=10cm
   \epsfysize=7cm
   \leavevmode
   \epsffile{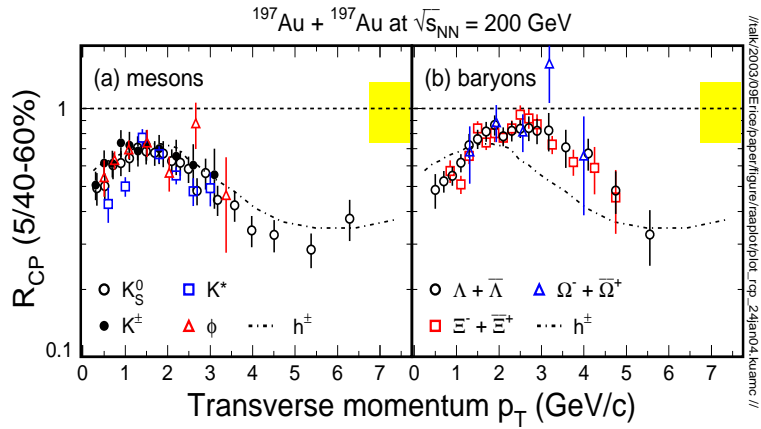} 
\caption[fig:rcp]{$R_{CP}$ of $K^{\pm}$, $K_{S}$, $K^{*}$, $\phi$,
$\Lambda$, $\Xi$ and $\Omega$ in comparison with charged hadron in dashed line. Distinct 
meson and baryon groups are observed.}
\label{fig:rcp}
\end{figure}

Figure~\ref{fig:rcp} shows the $R_{CP}$ of $K^{\pm}$, $K_{S}$, $K^{*}$, $\phi$, 
$\Lambda $, $\Xi $ and $\Omega $ as a function of $p_{T}$ from Au+Au collisions at 
$\sqrt{s_{NN}}=200$ GeV measured by the STAR collaboration, where the $R_{CP}$ ratio
is derived from the most central $5\%$ to the peripheral $40-60\%$ collision centralities. 
The dashed line is the 
$R_{CP}$ of charged hadrons for reference. In the low $p_{T}$ region soft
particle production is dominated by the number of participant scaling.
In the intermediate 
$p_{T}$ region of 2 to 5.5 GeV/c the $p_{T}$ dependence of $R_{CP}$ falls
into two groups, one for mesons and one for baryons. Despite the large mass
differences between $K_{S}$ and $K^{\ast }/\phi $, and between $\Lambda $ and 
$\Xi $ little difference among the mesons and among the baryons has been
observed within statistical errors. 
Particle dependence in the nuclear modification factor disappears only above a
$p_{T}$ of 6 GeV/c,
consistent with the expectation from conventional fragmentation processes. The
unique meson and baryon dependence in the intermediate $p_{T}$ region
indicates the onset of a production dynamics very different from both
fragmentation at high $p_{T}$ and hydrodynamic behavior at low 
$p_{T}$.

\begin{figure}[htb]
   \centering
   \epsfxsize=10cm
   \epsfysize=4.5cm
   \leavevmode
   \epsffile{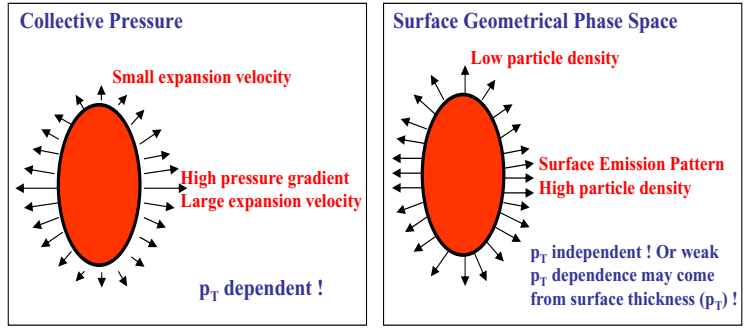} 
\caption[schem]{Schematic diagram to show two dynamical origins of angular anisotropy, one from
hydrodynamical expansion and the other from surface geometrical phase space.}
\label{schem}
\end{figure}

The azimuthal angular distribution can be described by a Fourier expansion 
\begin{equation*}
\frac{d^{2}n}{p_{T}dp_{T}d\phi }\propto (1+2\sum_{n}v_{n}\cos n(\phi -\Psi _{R}))
\end{equation*}
where $\phi$ is the azimuthal angle of the particle, $\Psi _{R}$ is the reaction plane angle and $v_{2}$ has been called 
elliptic flow~\cite{Sorge}. Theoretical calculations indicate that 
$v_{2}$ is generated mostly at the early stage of the nuclear collision.
There are two dynamical mechanisms responsible for $v_{2}$: hydrodynamic and
geometrical phase space of emitting source. In a non-central nucleus-nucleus collision the
overlapping participants form an almond shaped particle emission source. 
The reaction plane is defined by the vectors x (impact parameter direction) and z (beam direction).
The pressure gradient is greater and particles would experience larger
hydrodynamic expansion along the short axis direction (in plane) than these
along the long axis (out plane), resulting an ellipsoid in transverse
momentum space in the final state. Particle production from a dense matter
could lead to a surface emission pattern with higher particle density in the
reaction plane than that out of the reaction plane. Figure~\ref{schem} shows a
schematic diagram for two scenarios of generating azimuthal
angular anisotropy $v_{2}$ in non-central collisions.

\begin{figure}[htb]
   \centering
   \epsfxsize=8cm
   \epsfysize=5.5cm
   \leavevmode
   \epsffile{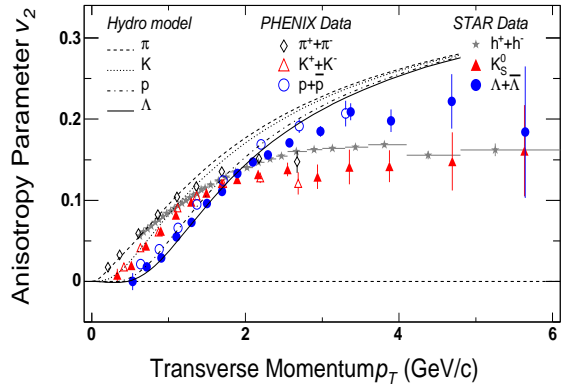} 
\caption[fig:v2]{Azimuthal angular anisotropy $v_2$ as a function of $p_{T}$ for identified particles. 
The hydrodynamic calculation is provided by P. Huovinen et al~\cite{huovinen}.}
\label{fig:v2}
\end{figure}

Figure~\ref{fig:v2} shows the $v_{2}$ as a function of $p_{T}$ for $\pi $, $K$, $p$, 
$\Lambda $ and $\Xi $ from PHENIX~\cite{phenix-v2} and STAR~\cite{star-v2} 
measurements. The angular anisotropy $v_{2}$ reveals
three salient features: 1) particles exhibit hydrodynamic behavior in the
low $p_{T}$ region -- a common expansion velocity may be established from the
pressure of the system and the heavier the particle the larger the momentum
from the hydrodynamic motion leading to a decreasing ordering of $v_{2}$
from $\pi $, to $K$ and $p$ for a given $p_{T}$; 2) $v_{2}$ values do not
depend strongly on $p_{T}$ at the intermediate $p_{T}$ region in contrast to
the strong $p_{T}$ dependence in the yield of particles; 3) the saturated $%
v_{2}$ values for baryons are higher than those for mesons and there is a
distinct grouping among mesons and baryons.

The absence of a strong $p_{T}$ dependence at the intermediate $p_{T}$
region is an intriguing phenomenon for the angular anisotropy of particle
emission. A very plausible scenario would be to relate the $v_{2}$ in this $%
p_{T}$ region with the geometrical shape of the emitting particle source. A
surface emission pattern from the almond shaped participant volume would
naturally lead to a saturation of $v_{2}$. A surface emission scenario is
possible if particles are produced from a dense bulk matter and the
hadronization duration is relatively short.

Parton energy loss in the dense medium created in nucleus-nucleus collisions
was proposed as a possible mechanism for generating an angular anisotropy $v_{2}
$. High energy partons are quenched inside the dense medium and lead to an
effective particle emission from a shell area of the participating volume~\cite{muller}.
Gyulassy and Vitev proposed the parton energy scenario as a mechanism to
explain the deviation of $v_{2}$ from hydrodynamic behavior at moderate $p_{T}$~\cite{Vitev}. 
While this scenario may be important for the considerable $v_{2}$
magnitude for charged hadrons at a $p_{T}$ greater than 6 GeV/c or so
measured by STAR~\cite{star-highpt-v2}, it cannot explain the particle dependence in both the
nuclear modification factor and the angular anisotropy $v_{2}$ at the
intermediate $p_{T}$ region. Within the parton energy loss scenario the
larger $v_{2}$ of baryons implies a higher energy loss than that of mesons;
the larger nuclear modification factor of baryons, however, is only
consistent with a smaller energy loss than that of mesons.

Shuryak~\cite{shuryak} has pointed out that the magnitude of the measured $v_{2}$ is
significantly larger than what can be accommodated based on particle
emission from a geometrical ellipsoid source within an energy loss scenario.
Using a more realistic Wood-Saxon description of the colliding nuclei for the
ellipsoid source the predicted theoretical $v_{2}$ is much smaller than that from
a hard-sphere model of the colliding nuclei, leading to a greater
discrepancy between the measurement and the theoretical expectation. 
The magnitude and the particle dependence of $v_{2}$ at the
intermediate $p_{T}$ region cannot have a dynamical origin either from hydrodynamic
flow or from parton energy loss alone.

\begin{figure}[htb]
   \centering
   \epsfxsize=8 cm
   \epsfysize=8 cm
   \leavevmode
   \epsffile{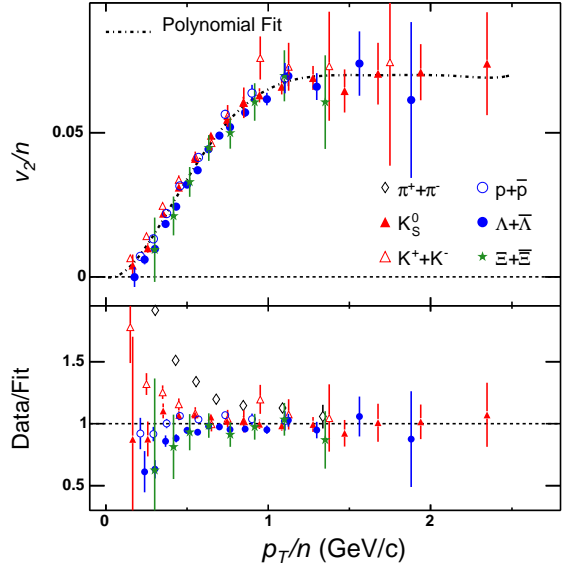} 
\caption[scale_v2]{Azimuthal angular anisotropy $v_2/n$ as a function of $p_{T}/n$ for identified
particles where n is the number of constituent quarks. The line is a fit to the data points excluding the
$\pi$ data and the bottom panel shows the deviation from the fit line.}
\label{scale_v2}
\end{figure}

An empirical constituent quark number ($n$) scaling has also been observed~\cite{sorensen}. 
Figure~\ref{scale_v2} presents
the $v_{2}/n$ as a function of $p_{T}/n$ for $\pi $, $K$, $p$, $\Lambda $
and $\Xi $ from Au+Au collisions, where the line is a fit to the data points
excluding the $\pi $ data. The bottom panel shows the ratio of data points
to the fit. At the intermediate $p_{T}$ region ( $0.6<$ $p_{T}/n$ $<2$
GeV/c) all meson and baryon data points fall onto an uniform curve. In the
lower $p_{T}$ region there is a particle dependent deviation from the fit
curve presumably due to the hydrodynamic flow as a function of mass of
particles. The $\pi $ data are significantly above the $v_{2}$ of other
mesons. The large fraction of resonance decay contribution to the $\pi $
yield is known to enhance the $v_{2}$ of $\pi $ at a given $p_{T}$~\cite{Greco,Dongx}.

The quark number scaling hints at dynamics for the hadron formation at the
intermediate $p_{T}$ which is very different from fragmentation at high 
$p_{T}$ and from thermal statistical descriptions at the low $p_{T}$ region. The
scaled azimuthal angular anisotropy ($v_{2}/n$) may be interpreted as the
constituent quark anisotropy just prior to the hadron formation. The essence
of this hadron formation dynamics is very different from the fragmentation
picture where the leading parton plays a dominant role in determining
properties of hadron. The measured features for intermediate $p_{T}$
hadrons produced at RHIC require that all quark ingredients (n=2 for mesons
and 3 for baryons) play an approximately equal role in hadron formation and that
the hadron properties are determined by the sum of partons.

Recent theoretical models of quark coalescence~\cite{voloshin,lin} and
recombination~\cite{Duke,Hwa} all have the essence of multi-parton dynamics for
hadron formation despite significant differences in details.
The recombination models provided a satisfactory description of
the particle yields, in particular, the large production of baryons in the
intermediate $p_{T}$ region. The formation of a dense partonic system
provides a parton density dependent increase in baryon yield as a function of collision
centrality through the coalescence mechanism. More importantly the particle
dependence in $v_{2}$ requires a $v_{2}$ distribution at the constituent
quark level. The establishment of the quark collectivity $v_{2}$ would be an
unambiguous signal for deconfinement in bulk partonic matter.

\section{Color Deconfined Matter and its QCD Properties at Hadronization}

The fact that for hadron formation at intermediate $p_{T}$ all quark
constituents must contribute almost equally to hadron properties does
not depend on details of theoretical models. Recombination~\cite{Duke,Hwa} or coalescence
models~\cite{voloshin,lin} provide a useful theoretical framework to derive quantitative quark
level properties. However, these models do not constitute unambiguous evidence
for deconfined partonic matter. The recombination model developed by Hwa and
Yang~\cite{Hwa-dAu}, for example, works for p+p and d+Au collisions as well. These
models provide an alternative hadronization scheme which, though different from
fragmentation, has been demonstrated to be more suitable for hadron
formation at the intermediate $p_{T}$ region from nucleus-nucleus collisions at
RHIC. The unique evidence for a color deconfined quark matter comes from the
fact that constituent quarks must have a collective $v_{2}$ distribution in
order to explain the measured particle dependence of azimuthal angular
anisotropy.

The particle dependences in the nuclear modification factor and in the
angular anisotropy $v_{2}$ have demonstrated that two distinct groups: mesons and baryons,
characterize the particle dependence at the intermediate $p_{T}$
region. The quark number scaling in the $v_{2}$ indicates that constituent
quarks have developed a collective angular anisotropy distribution $v_{2}$
in transverse momentum space by the time of hadronization and hadrons are
formed through coalescence or recombination of constituent quarks. The
dominant degrees of freedom at hadronization seem to be in the
constituent quarks. Based on nuclear parton distribution functions and quark
versus gluon interaction cross sections, the initial conditions of
nucleus-nucleus collisions at mid-rapidity at RHIC are expected to be dominated by gluon
degrees of freedom. We do not know how the colliding system
evolves from the initially gluon-dominated matter to a constituent quark
dominated system just prior to hadronization. The dominance of the
constituent quark degrees of freedom and the collective $v_{2}$ distribution
at the constituent quark level are the most direct experimental evidence for
a color deconfined matter, despite the lack of our understanding on the
dynamics of nuclear matter evolution. The saturation of $v_{2}$ can be
related to the geometrical shape of the particle emitting source at the time
of hadron formation, which supports the notion of bulk partonic matter of
constituent quark degrees of freedom.

The emerging physical picture implies that constituent quarks are the dominant degrees of freedom
at the boundary of quark-hadron transition. The
gluon degrees of freedom are not explicitly manifested in the hadron formation
or in characterizing the hadron properties. It appears that the nuclear matter
created in nucleus-nucleus collisions at RHIC would evolve naturally from
a partonic gluon-dominated initial state towards a constituent quark or
quasi-hadron state at the time of hadron formation. These experimental
observations may provide useful insights on the QCD properties of the quark
matter near $T_{c}$. Recent Lattice QCD (LQCD) calculations indicate that
spectral functions of pseudoscalar and vector mesons have non-trivial shapes
at a temperature above the critical $T_{c}$~\cite{Pete}. In
particular, heavy quarkonia such as J/$\psi $ may survive at a temperature
above $1.6T_{c}$~\cite{Hatsuda,Degal}. 
Other theoretical calculations, for example reference~\cite{brown}, have also invoked the notion
of quasi-hadrons to describe properties of the dense matter created at RHIC.
It is a critical step to firmly connect the experimental insights on the
properties of the quark matter at the boundary of hadronization with the
LQCD calculations. The disappearance of the gluon degrees of freedom from
the initial state and the emergence of constituent quarks at
hadronization are some of the critical conceptual questions to be addressed.
These questions will have a direct bearing on the nature of the quark hadron
phase transition for which there is as yet no direct experimental evidence.

\section{Charm Production Cross Section at RHIC}

Heavy quarks are produced mostly in initial parton scattering or during the
very high temperature phase of the collision. Therefore, heavy quark
measurement can probe the initial parton flux, the dynamical evolution and the quark
energy loss in dense medium. If heavy quarks are found to participate in the
collective motion of the medium (radial or elliptic flow), this will lend
further confirmation for parton collectivity. Features of charm
meson production will also bear signatures of the partonic matter at the
phase boundary. Both the STAR and PHENIX collaborations at RHIC have been
pursuing vigorous heavy quark physics programs both in analyses of current data and in 
future detector upgrade
plans to provide better capabilities for heavy quark measurements.

\begin{figure}[htb]
   \centering
   \epsfxsize=8cm
   \epsfysize=6.0cm
   \leavevmode
   \epsffile{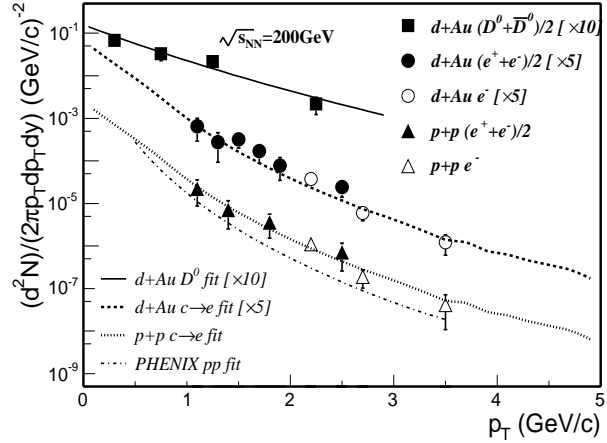} 
\caption[pt_ele]{$p_{T}$ distribution of $D^{0}$ mesons and non-photonic electrons from semi-leptonic 
decays of charm mesons.}
\label{pt_ele}
\end{figure}

The total charm quark pair production cross section ($\sigma _{c\overline{c}}$) 
is an important constraint on the collision dynamics and the heavy quark
evolution. Both STAR and PHENIX have presented results on the 
$\sigma _{c\overline{c}}$ measurement of p+p collisions from charm semi-leptonic
decays. In addition, STAR has also derived an equivalent $\sigma _{c\overline{c}}
$ for p+p collisions based on direct reconstruction of
hadronic decays of charm mesons from d+Au collisions. Figure~\ref{pt_ele} shows the $p_{T}$ spectra of
electron and $D^{0}$ from STAR. The PHENIX preliminary non-photonic electron data are represented by
the fitted line with a reported $\sigma _{c\overline{c}}$ measurement of $709\pm 85$%
(stat)$+332-281$(sys) $\mu b$~\cite{phenix-e}. 
STAR has measured a cross section of $1300\pm
200\pm 400$ $\mu b$ and a mean transverse momentum for $D^{0}$ of $1.32\pm
0.08$ GeV/c from direct $D^{0}$ reconstruction~\cite{star-e}. 
A next-to-leading order pQCD
calculation of the charm quark production cross section~\cite{vogt} has yielded 300 to 450 $\mu b$,
significantly below the STAR measurement and at the lower end of the PHENIX
range of uncertainty.

Several comments on the cross section measurements are in order. When
measuring the charm cross section through semi-leptonic decay, the quality of the electron data
below a $p_T$ of 1 GeV/c or so
is significantly deteriorated because of a large combinatorial background.
An electron of $p_{T}$ $\sim 1$ GeV/c typically comes from the decay of a 
$D$ meson with $p_{T}$ $\sim 2$ GeV/c, which is significantly beyond
the average $p_{T}$ of $1.32\pm 0.08$ GeV/c reported by STAR. Therefore, one
has to extrapolate to the low $p_{T}$ region by over a factor of two to
obtain the total charm cross section. Such an extrapolation is often model
dependent and has a large uncertainty. The semi-leptonic decay branching
ratios for $D^{0}$, $D^{\ast }$, $D^{\pm }$ and $D_{S}$ are different. The
electron yield from decays of these $D$ mesons depends on the relative yield
which is another important contribution to the uncertainties of the charm
production cross section derived from electron measurement. The direct
reconstruction of the $D$ decay kinematics provides a broad coverage of 
$p_{T}$ and does not suffer from the same uncertainties as the leptonic decay
electron measurement. However, present STAR measurements of $D$ meson yields using
event-mixing methods from TPC tracks suffer from limited statistics. A future
vertex detector upgrade capable of measuring the $D$ decay vertex
displacement is essential for both STAR and PHENIX heavy flavor physics
programs.

\begin{figure}[htb]
   \centering
   \epsfxsize=8cm
   \epsfysize=6.0cm
   \leavevmode
   \epsffile{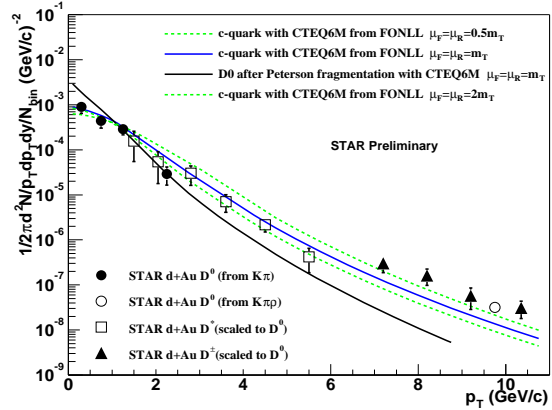} 
\caption[D_pt]{STAR preliminary measurement of $p_{T}$ distribution of $D$ mesons from d+Au collisions normalized 
by the number of binary collisions. The shape of the $p_{T}$ distribution is compared to pQCD FONLL calculations.}
\label{D_pt}
\end{figure}

Figure~\ref{D_pt} presents the STAR preliminary transverse momentum spectrum of $D$
mesons from d+Au collisions normalized by the number of binary collisions,
where the $p_{T}$ shapes of $D^{\ast }$ and $D^{\pm }$ are assumed to be the
same as that of $D^{0}$~\cite{atai}. The shape of the $p_{T}$ distribution coincides
with the bare charm quark $p_{T}$ distribution from the Fixed-Order-Next-Leading-Log (FONLL)
pQCD calculation from M. Cacciari et al~\cite{fonll}. If a fragmentation function such as the
Lund fragmentation scheme~\cite{pythia} or the Peterson function~\cite{peterson} is introduced
for $D$ meson production, the resulting $p_{T}$ distribution will be
significantly below the measurement at the high $p_{T}$ region. This
observation raises an outstanding question regarding the $p_{T}$
distribution and the formation mechanism of $D$ mesons in hadro-production.
Recently
a $k_{T}$ factorization scheme has been found to significantly change the $D$ meson $p_T$
distribution from nuclear collisions as well~\cite{Tuchin}.

The fact that the $D$ meson $p_{T}$ distribution can be better described by
the $p_{T}$ of bare charm quarks from pQCD calculations has been observed in
previous fixed target experiments~\cite{FONLL-1}. With a fragmentation function such as
the Peterson function the $D$ meson $p_{T}$ distribution is too steep to
explain the measured $p_{T}$ distribution. In order to match the calculation
with the data one has to introduce a $k_{T}$ kick to the parton
distribution. The scale of $<k_{T}^{2}>$ is on the order of 1 (GeV/c)$^{2}$,
much larger than the typical $\Lambda _{QCD}$ scale for strong
interaction. Furthermore, the Feynman $x_{F}$ of the $D$ meson distribution
was also found to coincide with the bare charm quark $x_{F}$ distribution~\cite{FONLL-1}. 
The fragmentation function will have a large impact on the $x_{F}$
distribution from bare charm quark to $D$ meson, which cannot be negated by
introducing any $k_{L}$ longitudinal boost of reasonable scale as in
the case for $k_{T}$ kick in the transverse momentum direction.

The transverse momentum distribution of particles produced at RHIC energies
are considerably flatter than those at lower energies. The $k_{T}$ kick
scheme does not change the shape of the $p_{T}$ distribution significantly.
The STAR measurement of the $D$ meson $p_{T}$ distribution suggests either that
the charm quark fragmentation may be close to a delta function or that a charm
formation mechanism such as the recombination model may be important in
hadro-production. The recombination model takes a charm quark and combines
it with a light quark, presumably of low $p_{T}$, to form a charmed meson.
Therefore, the $p_{T}$ of the meson is not significantly different from that
of the bare charm quark. Measurements of charmed meson production provide
unique probes for the hadron formation dynamics
and for the transport dynamics of heavy quarks in the dense nuclear medium 
produced at RHIC energies.

\section{Recombination Mechanism and Heavy Quark Flow and Energy Loss}

R.J. Fries {\it et al}~\cite{Duke} has argued that recombination dominates over the
fragmentation production mechanism when the parton spectrum is a thermal
distribution. The resulting hadron momentum distribution exhibits the same
thermal distribution as the constituents. Only at sufficiently large $p_{T}$, 
where the underlying parton distribution is a power-law, does
fragmentation overtake recombination. The recombination process is
particularly effective for baryon production which recombines three readily
available quarks while in fragmentation process the three quark production
is suppressed. Since the low $p_T$ distributions of hadrons from
p+p, d+A and A+A collisions at high energies such as RHIC are not drastically
different quanlitatively, the argument by R.J. Fries et al would imply that the recombination
process should play a unique role at the intermediate $p_{T}$ region for all these
collisions if we interpret the soft particle production as from an
underlying thermal parton distribution. Recombination model calculations by Hwa and
Yang can fit Au+Au and d+Au data from RHIC reasonably well~\cite{Hwa-dAu}. The fact
that phenomenologically an underlying parton distribution is used for the
recombination model does not necessarily mean the formation of partonic
matter.

The experimental evidence of bulk partonic matter formation and its
collectivity comes from the elliptic flow $v_{2}$ of partons based on
constituent quark scaling in the $v_{2}$ measurement of mesons and baryons.
Therefore, one critical test of the idea is to measure many more particles
and resonances over a much broader $p_{T}$ region. In particular, we need to
push the $p_{T}$ coverage above 6 GeV/c or so to reach the fragmentation
region. The meson and baryon difference in the nuclear modification factor
and the angular anisotropy $v_{2}$ is expected to vanish in the
fragmentation region. 

STAR preliminary measurement of $\Xi $ and its
comparison with $\Lambda $ indicated that the strange quark dynamics in
nuclear modification factor and the $v_{2}$ at the intermediate $p_{T}$
region is similar to these of light quarks: the constituent strange quarks
have a similar $v_{2}$ as light up and down quarks~\cite{schweda}. The measurement of
nuclear modification factor and $v_{2}$ of $\phi $ meson and $\Omega $
baryon covering the full $p_{T}$ range from recombination to fragmentation
region in nucleus-nucleus collisions at RHIC will further test the strange
quark dynamics in the hadronization of bulk partonic matter. Possible
variation may come from the particle formation differences because the $\phi $ 
meson is a vector meson and $\Omega $ belongs to decuplet while what have
been measured are for pseudo scalar mesons and octet baryons. Resonance
measurement has also been proposed to test the production dynamics~\cite{Nonaka}.

\begin{figure}[htb]
   \centering
   \epsfxsize=7.5cm
   \epsfysize=6.5cm
   \leavevmode
   \epsffile{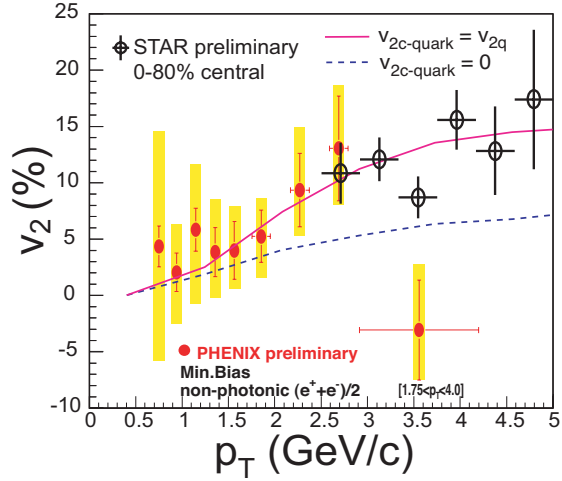} 
\caption[ele-v2] {Preliminary measurement of non-photonic electron $v_2$ as a function 
of $p_T$ from PHENIX and STAR. The curves are from a calculation by V. Greco et al~\cite{Rapp}.}
\label{ele-v2}
\end{figure}

Charm quark transport dynamics in dense nuclear medium will provide unique
probes to the QCD properties of the medium. 
If the initial temperature is very high, $T\sim 500$ 
MeV or higher, the yield of total charm quarks can also be increased through
thermal gluon-gluon scatterings. 
Possible suppression of charm
mesons at high $p_{T}$ will test the energy loss dynamics of charm quark
propagation in a QCD medium. Theoretical calculations have predicted a reduced
medium induced energy loss for heavy quarks and the high $p_{T}$ suppression
of charmed hadrons should not be as strong as light hadrons~\cite{Kharzeev,MG}. 
Observation
of heavy quark hydrodynamic flow would indicate that heavy quarks, once
created in the initial state, must have participated in the partonic
hydrodynamic evolution over a sufficiently long period of time to reach a
substantial flow magnitude. This would be a unique probe for the early stage
of a partonic phase~\cite{Rapp}. Figure~\ref{ele-v2} shows preliminary
STAR~\cite{star-e-v2} and PHENIX~\cite{phenix-e-v2}  measurements of the $v_{2}$ for 
electrons from charm leptonic
decays, which have been demonstrated to be closely correlated with charmed
meson $v_{2}$~\cite{Dongx}.

\section{Summary}

Characteristics in hadrons below a  $p_{T}$ of 5 GeV/c or so from
nucleus-nucleus collisions at RHIC have shown distinct features that are
drastically different from fragmentation processes in elementary collisions.
A salient feature of meson and baryon dependence has been observed in the
nuclear modification factor and the angular anisotropy $v_{2}$ of $\pi $, 
$K^{\pm }$, $K_{S}$, $K^{\ast }$, $p$, $\Lambda $ and $\Xi $ particles at
the intermediate $p_{T}$ of 2-5 GeV/c. A constituent quark number scaling has been observed
for the $v_{2}$ measurement. These experimental measurements suggest that at
the hadronization moment the effective degrees of freedom are the
constituent quarks; the constituent quarks have developed a collective $v_{2}$ 
as a function of $p_{T}$; and the hadron formation at the intermediate 
$p_{T}$ is likely through multi-parton dynamics such as 
recombination or coalescence process.

The physical picture emerging from the experimental measurements complements
the Lattice QCD results. Spectral function calculations have indicated that
hadrons, particularly heavy quarkonia, do not melt completely at critical
temperature. It appears plausible that the constituent quark degrees of
freedom or quasi-particles play a dominant role at the hadronization of bulk
partonic matter though further confirmation of the picture from LQCD is
needed. Despite intriguing experimental observations of hadronization from a
deconfined bulk partonic matter, signatures for a quark-hadron phase
transition remain elusive.

Heavy quark production and its transport dynamics in dense nuclear medium
probe QCD properties of the matter. The charm quark flow measurement will
provide a significant insight on recombination or coalescence hadronization
mechanism and partonic collectivity of the dense matter. Future detector
upgrades from STAR and PHENIX will greatly enhance their heavy quark
measurement capabilities at RHIC.

\section{Acknowledgment}
We thank An Tai, Hui Long, Paul Sorensen, Xin Dong, Frank Laue, Zhangbu Xu, Nu Xu, 
Jan Rafelski and Charles Whitten Jr. for many stimulating discussions on physics topics in this article
and for their help on the manuscript.

\end{document}